\documentclass{elsart}

\usepackage{amsmath}
\usepackage{psfig}
\usepackage{fancyhdr}
\usepackage{rotate}
\usepackage{rotating}
\usepackage{dcolumn}
\usepackage{bm}
\usepackage{subfigure}
\usepackage{epsfig}
\usepackage{natbib}
\usepackage{graphicx}
\usepackage{color}
\usepackage{latexsym}
\usepackage{amsmath}

\newcommand{\vect}[1]{\mbox{\boldmath$#1$\unboldmath}}

\newcommand{\M}{mol~l$^{-1}$}

\newcommand{\degc}{$^\circ$C}
\newcommand{\ie}{{\it i.e.}}
\newcommand{\eg}{{\it e.g.}}

\newcommand{\bsl}{{\ell}}

\newcommand{\Pe}{\mbox{Pe}}

\newcommand{\kb}{k}
\newcommand{\ka}{\kappa a}

\begin{document}

\begin{frontmatter}
  
  \title{Electric-field-enhanced transport in polyacrylamide hydrogel
    nano-composites}

\author{Reghan J. Hill}

\address{Department of Chemical Engineering and McGill Institute for Advanced
  Materials, McGill University, Montreal, Quebec, H3A 2B2, CANADA}

  \begin{abstract}
    Electroosmotic pumping through uncharged hydrogels can be achieved by
    embedding the polymer network with charged colloidal inclusions.
    \cite{Matos:2006} recently used the concept to enhance the
    diffusion-limited flux of uncharged molecules across polyacrylamide
    hydrogel membraness for the purpose of improving the performance of
    biosensors. This paper seeks to link their reported macroscale diagnostics
    to physicochemical characteristics of the composite microstructure. The
    experiments are characterized by a Debye screening length that is much
    larger than the radius of the silica nanoinclusions and the Brinkman
    screening length of the polymer skeleton. Accordingly, closed-form
    expressions for the incremental pore mobility are derived, and these are
    evaluated by comparison with numerically exact solutions of the full
    electrokinetic model. A mathematical model for the bulk electroosmotically
    enhanced tracer flux is proposed, which is combined with the
    electrokinetic model to ascertain the electroosmotic pumping velocity from
    measured flux enhancements. Because the experiments are performed with a
    known current density, but unknown bulk conductivity and electric field
    strength, theoretical estimates of the bulk electrical conductivity are
    adopted. These account for nano-particle polarization, added counterions,
    and non-specific adsorption. Theoretical predictions of the flux
    enhancement, achieved without any fitting parameters, are within a factor
    of two of the experiments. Alternatively, if the Brinkman screening length
    of the polymer skeleton is treated as a fitting parameter, then the
    best-fit values are bounded by the range 0.9--1.6~nm, depending on the
    inclusion size and volume fraction. Independent pressure-driven flow
    experiments reported in the literature for polyacrylamide gels without
    inclusions suggest 0.4 or 0.8~nm. The comparison can be improved by
    allowing for hindered ion migration, while uncertainties regarding the
    inclusion surface charge are demonstrated to have a negligible influence
    on the electroosmotic flow. Finally, and perhaps most importantly,
    anomalous variations in the flux enhancement with particle size and volume
    fraction can be rationalized at present only by acknowledging that
    particle-particle and particle-polymer interactions increase the effective
    permeability of the hydrogel skeleton. This bears similarities to the
    increase in polymer free volume that accompanies the addition of silica
    nanoparticles to certain polymeric membranes.
\end{abstract}

\end{frontmatter}  

  \section{Introduction} \label{sec:introduction}
  
  Electric fields are widely used to probe the physicochemical characteristics
  of dispersed colloidal particles. Microelectrophoresis, capillary
  electrophoresis and gel electrophoresis, for example, are routinely used to
  characterize and sort DNA fragments, proteins, synthetic and biological
  macromolecules, and supramolecular aggregates. In all these applications,
  the electric field induces translation (electrophoresis) of charged
  particles, with the electrophoretic velocity reflecting a balance of
  electrical, hydrodynamic and steric interaction forces, which, in turn, are
  related to the particle size and charge, and properties of the dispersing
  medium.
  
  \cite{Matos:2006} recently proposed a novel means of promoting the transfer
  of uncharged molecules across uncharged polyacrylamide hydrogels. By
  embedding the polymer skeleton with immobilized silica nanoparticles, they
  endowed the matrix with a fixed charge, and demonstrated that the
  application of an electric field enhanced the otherwise diffusion-limited
  flux of uncharged tracer across the membrane. Despite the possibility of
  promoting unwanted electrostatic interactions with charged molecules, the
  ability to control the effective charge density by varying the concentration
  of immobilized inclusions offers several advantages over chemical and/or
  physicochemical modifications of the polymer architecture alone. Specific
  interactions may also be imparted by decorating the inclusions with
  functional moieties, such as in affinity chromotography.
  
  As expected from studies involving electroosmotic microfluidic
  pumping~\citep{Yao:2003a,Yao:2003b}, enhanced transport in immobilized cell
  culture~\citep{Chang:1995} and electroosmotic
  dewatering~\citep{Yukawa:1976}, the origin of the electric-field induced
  flux enhancement in Matos~\etal's experiments was qualitatively attributed
  to electroosmotic flow originating from the diffuse layers of charge that
  envelop the silica inclusions.  Moreover, the microstructure of the
  composites synthesized by \citeauthor{Matos:2006} has potential to serve as
  a model system with which to test electrokinetic theory that predicts, from
  first principles, the average electroosmotic pumping velocity, electrical
  conductivity, and effective ion diffusion coefficients in this novel class
  of nano-structured composites~\citep{Hill:2006a,Hill:2006b}. Indeed, the
  purpose of this paper is to provide a quantitative theoretical
  interpretation of the available experiments and, in turn, link macroscale
  diagnostics to characteristics of the microstructure.
  
  The paper is set out as follows. First, a detailed description of
  Matos~\etal's experiments is provided in \S\ref{sec:exp}. This identifies
  the relevant parameters and highlights the available diagnostics.
  Section~\ref{sec:theory} presents various aspects of the theory. In
  particular, \S\S\ref{sec:emodel} describes the electrokinetic model;
  \S\S\ref{sec:flux} derives the means of linking Matos~\etal's flux
  enhancement ratio to the electroosmotic pumping velocity;
  \S\S\ref{sec:polar} addresses the bulk electrical conductivity of the
  composite membranes; and \S\S\ref{sec:mobilityapprox} evaluates---by
  comparison with numerical solutions of the full electrokinetic model---a
  convenient closed-form approximation for the incremental pore mobility. The
  results are presented in \S\ref{sec:results}. Here, literature values of the
  permeability of gels without inclusions---obtained from pressure-driven
  flows---are compared to values inferred from the electrokinetic
  interpretation of the experiments. Certain anomalies are discussed in
  \S\ref{sec:discussion}, and the paper concludes with a summary
  (\S\ref{sec:summary}).

\section{Experimental} \label{sec:exp}

The hydrogel nanocomposite membranes synthesized in Matos~\etal's experiments
comprise a polyacrylamide hydrogel matrix with silica nanosphere inclusions.
The hydrogels were synthesized with a polymer mass fraction $c = 0.15$ (mass
of polymer to mass of water). After polymerization and cross-linking, however,
the matrix was reported to swell, yielding a final polymer mass fraction $c
\approx 0.13$. The microstructure of the membranes is characterized by the
radius of the inclusions ($a \approx 3.5$ or $7.5$~nm) and the inclusion
volume fraction ($\phi \approx 0.02$ or $0.04$).

\begin{figure}
  \begin{center}
    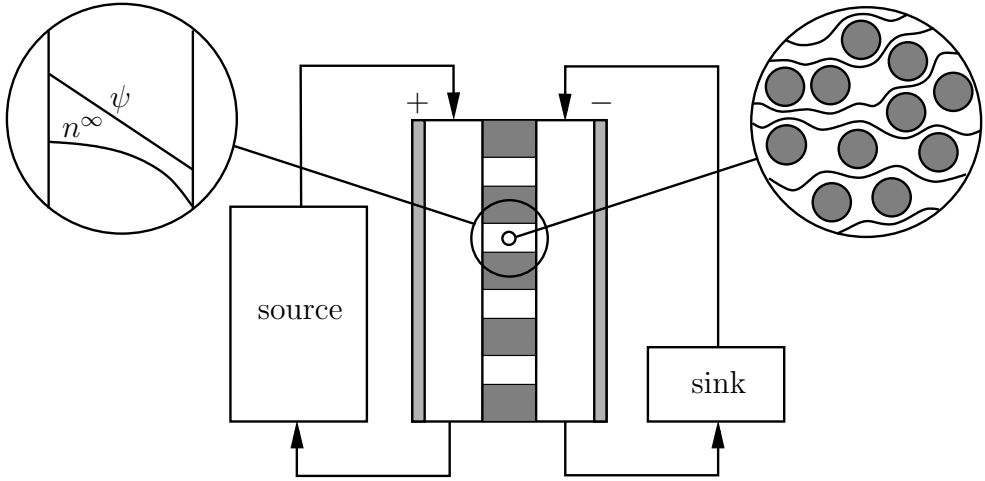
  \end{center}
  \caption{\label{fig:schematic} Schematic of the experimental
    apparatus used by \cite{Matos:2006} to measure the electric-field
    enhanced flux of a neutral tracer across polyacrylamide
    hydrogel-silica nanocomposite membranes. The immobilized silica
    nanoinclusions endow the microstructure with a negative charge,
    which, with an electric field $E = - \mbox{d} \psi / \mbox{d}x$
    between the electrodes, drives electroosmotic flow from the anode
    ($+$, left) toward the cathode ($-$, right). The flow produces a
    convective enhancement of the otherwise purely diffusive flux of
    the neutral tracer, whose concentration is denoted $n^\infty(x)$,
    from the source resevoir to the sink resevoir. See the text and
    Matos~\etal's paper for details.}
\end{figure}

A schematic illustration of Matos~\etal's apparatus is presented in
figure~\ref{fig:schematic}. The membrane comprises 29 cylindrical
sub-membranes, each with a diameter 2.68~mm (giving a total cross-sectional
area $A \approx 1.64$~cm$^2$) and thickness $L \approx 1.68$~mm. Matos~\etal \ 
measured the diffusion coefficient of the tracer (amino-methylcoumarin) in gel
membranes with and without inclusions, establishing a value $D \approx 3.3
\times 10^{-10}$~m$^2$s$^{-1}$. They did not discern the $O(\phi) \ll 1$
influence of the impenetrable inclusions on the effective diffusion
coefficient.

The experiments were initiated by establishing a quasi-steady
diffusive flux of uncharged tracer across each membrane before
applying a potential difference to the electrodes. The electric field
was applied after 30~min, which is short compared to the
characteristic diffusion time scale $L^2 / D \approx
140$~min. However, it is evident from the experiments and, indeed, the
exact solution of the transient diffusion problem~\citep{Matos:2006},
that the fastest transients decay in about 30~min.

The gold electrodes, which present a circular surface area on each
side of the membrane, have a diameter 1.9~cm and are separated by a
distance 6~mm. In the experiments considered in this paper, the
differential voltage was controlled to maintain a constant electrical
current $i \approx 3$~mA.

The experiments were performed with the source and the sink reservoirs
initially filled with either 1~m\M KNO$_3$ electrolyte or deionized
(DI) water. Only the source reservoir contained tracer (with
concentration 0.1~m\M) at the beginning of the experiments, and the
electrolytes on each side of the membrane were recirculated at a rate
$Q \approx 8$~cm$^3$~min$^{-1}$, with respective source and sink
reservoir volumes $V_1 \approx 50$~cm$^3$ and $V_2 \approx 10$~cm$^3$.

The residence time of fluid in the two gaps between the membrane and
the electrode is $\delta t \approx 3.6$~s, so the thickness of a
diffusive boundary layer at the membrane surface is $\delta l \approx
\sqrt{\delta t D} \approx 35~\mu$m. Equating the diffusive flux across
the membrane $D \Delta n^\infty / L$ to the flux across the boundary
layer $D \delta n^\infty / \delta l$ indicates that $\delta n^\infty /
\Delta n^\infty = \delta l / L \sim 10^{-2}$, so the tracer
concentration at the edges of the membrane is approximately equal to
the bulk concentration in the respective reservoir. Note that the
hold-up time of the tracer in the reservoirs is less than $V_1 / Q
\approx 6$~min, which is short compared to the $\sim V_2 L / (A D)$
characteristic time scale for the (diffusive) tracer flux to change
the reservoir concentration by an $O(1)$ amount. It follows that the
membrane and reservoir dynamics are quasi-steady.

The tracer (amino-methylcoumarin) concentration was measured with a
fluorescence spectrophotometer that continuously sampled fluid in the
sink reservoir circuit. \citeauthor{Matos:2006} also measured the pH
in both reservoirs, showing that the source and sink reservoirs,
respectively, attain acidic and basic pH with the application of an
electric field.

Note that the tracer has an acid dissociation constant $K_a =
10^{-5}$~\M, so the fraction of it in a protonated state, $1 / (1 +
K_a / [\mbox{H$_3$0$^+$}])$, is significant only when $\mbox{pH} <
5$. These conditions prevail in the source reservoir and, possibly,
throughout the composite membrane, suggesting that electrophoresis of
the tracer should contribute to the net electric-field-induced flux
enhancement.  From the diffusion coefficient $D \approx 3.3 \times
10^{-10}$~m$^2$~s$^{-1}$, the electrophoretic mobility is (to a first
approximation) $e D / (\kb T) \approx
1.1$~($\mu$m~s$^{-1}$)/(V~cm$^{-1}$). Therefore, with a representative
electric field $E \sim 10$~V~cm$^{-1}$, the electrophoretic velocity
of the tracer is $U_t \sim 10$~$\mu$m~s$^{-1}$, which corresponds to
$\Pe = U_t L / D \sim 55$. This is significantly greater than the
values ascertained from electroosmotic pumping, suggesting that the
flux of protonated tracer might be dominated by
electrophoresis. However, the experimental evidence is that the flux
is dominated by electro-osmosis~\citep{Matos:2006}. It follows that
$\mbox{pH} > 5$ in a significant portion of the composite hydrogel
membranes.

Finally, from the tracer concentration time series,
\citeauthor{Matos:2006} reported averaged flux enhancement ratios,
which they defined as the ratio of the tracer concentration in the
sink reservoir after 300~min (with an electric field applied after the
first 30~min) to the concentration from an experiment without an
electric field. They appropriately termed this ratio an {\em
enhancement factor}, whereas the analysis undertaken in this paper
approximates it as a {\em flux enhancement factor}, $J / J_0$. The
ratio of the fluxes, which undoubtedly exhibit weak transients, are
therefore assumed to be constant in this work. It is evident that
individual experiments reflect significant, non-systematic statistical
variations, so the analysis below relies entirely on five averaged
values of the reported enhancement factor, each with various particle
radius $a$ and volume fraction $\phi$, and electrolyte (1~m\M KNO$_3$
or deionized water), but fixed membrane geometry and electrical
current.

\section{Theory} \label{sec:theory}

\subsection{Electrokinetic model and electoroosmotic flow in uniform composites} \label{sec:emodel}

The electrokinetic transport model adopted in this work is described
in considerable detail elsewhere~\citep{Hill:2006b}. Briefly, it
comprises the non-linear Poisson Boltzman equation for electrostatics;
electrolyte conservation equations involving diffusion,
electromigration and convection; and Brinkman's equations (with an
electrical body force) for solvent mass and momentum
conservation~\citep{Brinkman:1947}. A solution of the equations, with
a single spherical charged, impenetrable colloid embedded in an
unbounded, rigid polymer gel yields the far-field ($r \rightarrow
\infty$) decays of the electrostatic potential $\psi$, ion
concentrations $n_j$ and fluid velocity $\vect{u}$. Of prime
importance in this paper are perturbations to equilibrium that arise
from a uniform electric field $\vect{E}$. In the far-field, these take
the following asymptotic forms:
\begin{eqnarray} \label{eqn:farfield1}
  \psi &\rightarrow& - \vect{E} \cdot \vect{r} + (\vect{E} \cdot
  \vect{e}_r) D^E r^{-2} \\ n_j &\rightarrow& n_j^\infty + (\vect{E}
  \cdot \vect{e}_r) C^E_j r^{-2} \\ \vect{u} &\rightarrow& - 2 C^E
  r^{-3} (\vect{E} \cdot \vect{e}_r) \vect{e}_r - C^E r^{-3} (\vect{E}
  \cdot \vect{e}_\theta) \vect{e}_\theta.
\end{eqnarray}
Note that $\vect{e}_r$ and $\vect{e}_\theta$ are radial and tangential
unit vectors, with the polar axis $\vect{e}_z$ being colinear with
$\vect{E}$. The scalar coefficients $D^E$, $C_j^E$ and $C^E \eta /
\bsl^2$ are, respectively, the dipole strengths of the electrostatic
potential, ion concentrations, and pressure perturbations induced by
the forcing $\vect{E}$. In general, these so-called asymptotic
coefficients are complicated functions of the particle radius and
$\zeta$-potential, $a$ and $\zeta$, the electrolyte viscosity and
Debye screening length, $\eta$ and $\kappa^{-1}$, and the Darcy
permeability of the polymer skeleton $\bsl^2$ (square of the Brinkman
screening length).

The single-particle problem leads to averaged equations that quantify
the bulk ion fluxes and fluid flow in a dilute ($\phi [1 +
(\ka)^{-1}]^3 \ll 1$) composite~\citep{Hill:2006b}. For the
statistically uniform composites addressed in this work, the pressure
and bulk ion concentration gradients are zero. Accordingly, with a
uniform electric field and unidirectional flow and electrolyte and
tracer fluxes, the average fluid velocity is
\begin{equation} \label{eqn:avemom}
  U = -\phi (3/a^3) E C^E,
\end{equation}
where $\phi \ll 1$ is the inclusion volume fraction, and the average
ion fluxes are
\begin{eqnarray}  \label{eqn:ionflux}
  J_j = n_j^\infty U + z_j e \frac{D_j}{\kb T} n_j^\infty E + \phi
  (3/a^3) z_j e \frac{D_j}{\kb T} n_j^\infty E D^E + \phi (3 / a^3)
  D_j E C_j^E,
\end{eqnarray}
where $z_j = \pm z$ ($j = 1,2$) are the electrolyte ion valences,
$D_j$ are the ion diffusion coefficients, $\kb T$ is the thermal
energy, and $e$ is the fundamental charge. In
Eqn.~(\ref{eqn:ionflux}), the first two terms are, respectively, the
convective and electromigrative fluxes that prevail in the absence of
inclusions; and the last two terms are (microscale) electromigrative
and diffusive contributions to the bulk flux arising from disturbances
of the inclusions.

\subsection{Uncharged tracer flux and flux enhancement} \label{sec:flux}

The flux of uncharged tracer is
\begin{eqnarray}  \label{eqn:tracerflux}
  J = n^\infty U - D B + \phi (3 / a^3) D C_3^B B,
\end{eqnarray}
where $C_3^{B} = (1/2) a^3$ and $B = \mbox{d} n^\infty / \mbox{d}x$ is
the bulk concentration gradient~\citep{Hill:2006a}.

Clearly, the electric-field-induced flow $U$ and electrolyte ion
fluxes $J_1$ and $J_2$ are decoupled from the tracer, but the tracer
flux is influenced by the electroosmotic flow. Therefore, the tracer
concentration satisfies the conservation equation
\begin{equation}
  \partial n^\infty / \partial t = - \partial J / \partial x,
\end{equation}
where
\begin{equation}
  J = n^\infty U - D [1 - (3/2) \phi]\partial n^\infty / \partial x.
\end{equation}

Under quasi-steady conditions ($\partial n^\infty / \partial t \approx
0$), which will be justified shortly, the flux may be written in terms
of the tracer concentration at the edges ($x=0$ and $L$) of the
membrane\footnote{The tracer concentration is $n = n^\infty(x=0)
e^{\Pe x / L} + (J / U) (1 - e^{\Pe x / L})$.}:
\begin{equation}
  J = U [n^\infty(x=L) - n^\infty(x=0) e^{\Pe}]/(1-e^{\Pe}).
\end{equation}
Here, the P{\'e}clet number is
\begin{equation} \label{eqn:peclet}
  \Pe = U L / D = \phi M E L / D
\end{equation} 
and
\begin{equation}
  M \equiv U / (E \phi) = - (3 / a^3) C^E[\ka, \zeta e/(\kb T), \kappa \bsl]
\end{equation}
is termed the {\em incremental pore mobility}.

Scaling arguments that also consider transients in the tracer
concentration in the (well mixed) source and sink indicate that a
quasi-steady approximation is reasonable when $V / (LA) \gg 1$ and
$\Pe \sim 1$; otherwise, $1 \ll \Pe \ll V / (LA)$ or $LA / V \ll \Pe
\ll 1$. Here, $V$ is the smallest of the source and sink volumes. For
Matos~\etal's experiments, $V \sim 10$~cm$^3$, $A \sim 1$~cm$^2$ and
$L \sim 0.1$~cm, so $V / (LA) \sim 10 \gg 1$, as required when $\Pe
\sim 1$.


With zero electric field ($\Pe = 0$), the tracer flux is
\begin{equation}
  J_0 = - D [n^\infty(x=L) - n^\infty(x=0)] / L,
\end{equation}
so the flux enhancement $J / J_0$ may be written
\begin{equation}
  J / J_0 = [1 + \epsilon (1 - e^{\Pe})] \Pe / (e^{\Pe} - 1),
\end{equation}
where $\epsilon = n^\infty(x=0) / [n^\infty(x=L) - n^\infty(x=0)]$.

If $J_0 > 0$ with $\epsilon \approx -1$, which corresponds to
diffusion from the source to the sink, with the tracer concentration
in the sink much less than in the source, it may be reasonable to
write
\begin{equation} \label{eqn:expmobility}
  J / J_0 \approx \Pe / (1 - e^{-\Pe}).
\end{equation}
Clearly, the flux is {\em enhanced} by an electric field that produces
an electroosmotic flow directed {\em down} the concentration gradient
($\Pe > 0$).

\subsection{Polarization, added counterions and non-specific adsorption} \label{sec:polar}

The ion mobilities in the theory above are unhindered by the polymer
skeleton when they are calculated from widely available limiting
conductivities. Therefore, the bulk electrical conductivity of the gel
in the absence of inclusions takes the same value as the electrolyte
in the absence of polymer. Furthermore, the experimentally determined
mobilities depend on a theoretical prediction of the bulk electrical
conductivity of the composite, taking into account concentration and
electrical polarization, added counterions, and non-specific
adsorption. Accordingly, the effective conductivity may be written
\begin{equation} \label{eqn:conductivity}
  K^e = K^\infty [1 + \phi(\Delta_p + \Delta_{ac} + \Delta_{nsa})],
\end{equation}
where the polarization increment $\Delta_p$ is obtained from the ion
fluxes above, giving
\begin{equation} \label{eqn:polarization}
  \Delta_p = (3/a^3) D^E e^2 / (\kb T) \sum_{j=1,2} z_j^2 D_j
  n_j^\infty / K^\infty + (3/a^3) e \sum_{j=1,2} z_j D_j C_j^E /
  K^\infty.
\end{equation}
This reflects the electric-field-induced electrostatic potential and
concentration dipole moments ($D^E$ and $D_j^E$) for a single
immobilized inclusion~\citep{Hill:2006b}. In contrast, the added
counterion and non-specific adsorption increments do not depend on
perturbations induced by the electric field; they can therefore be
calculated from knowledge of only the equilibrium base
state~\citep{Saville:1983,Hill:2003b}, as parameterized by $\zeta e /
(\kb T)$ and $\ka$.

\begin{figure}
  \begin{center}
    \includegraphics[width=8cm]{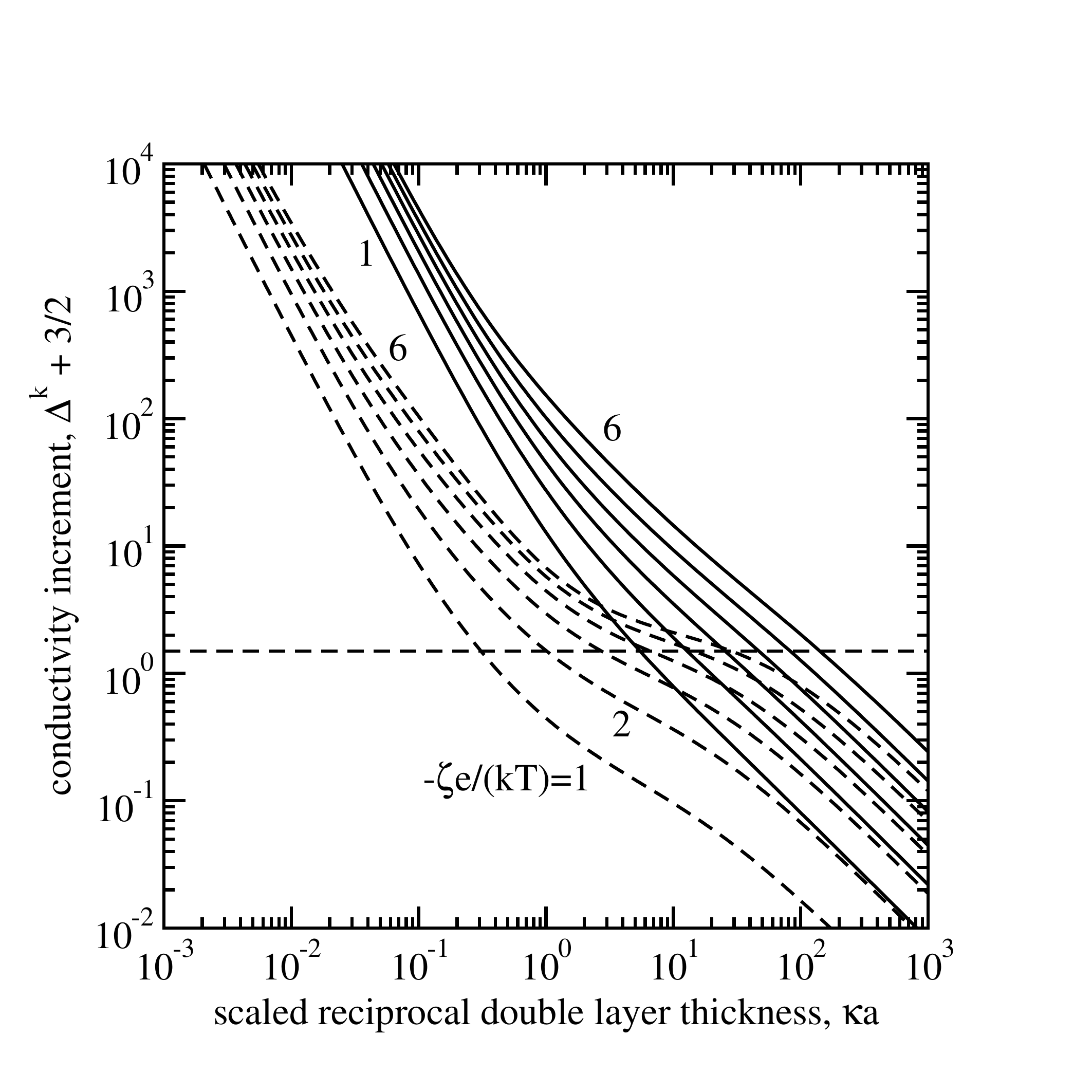}
  \end{center}
  \caption{\label{fig:condincrement} The conductivity increment ($3/2$
    has been added to permit a logarithmically scaled ordinate)
    $\Delta^k = \Delta_p + \Delta_{ac} + \Delta_{nsa}$ (solid lines)
    as a function of the reciprocal double layer thickness $\ka$ for
    various (scaled) $\zeta$-potentials $-\zeta e/(\kb T) = 1$, $2$,
    $3$, ..., $6$: aqueous KNO$_3$ at $T = 25$\degc; $a = 10$~nm;
    $\bsl = 1$~nm. Dashed lines are the contributions from
    electrostatic and concentration polarization, $\Delta_p$; these
    are practically independent of the Brinkman screening length
    $\bsl$. The horizontal line at $3/2$ denotes where $\Delta^k =
    0$.}
\end{figure}

The total conductivity increment $\Delta_p + \Delta_{ac} +
\Delta_{nsa}$ is plotted as a function of $\ka$ for various
$\zeta$-potentials in figure~\ref{fig:condincrement} (solid
lines). Recall, the added counterion and non-specific adsorption
contribution $\Delta_{ac} + \Delta_{nsa}$ is an $O(\phi)$ contribution
to the equilibrium base state, whereas the polarization contribution
$\Delta_p$ (dashed lines) is an $O(\phi)$ contribution to the
perturbed equilibrium. At low ionic strengths (small $\ka$),
$\Delta_p$ is smaller than for dispersions where the particles undergo
electrophoresis in a Newtonian
electrolyte~\citep{Hill:2006b}. Therefore, when the particles are
immobilized in a polymer network, the added counterion and
non-specific adsorption contribution plays a significant role.

Note that the calculations presented below for the experiments with DI
water as the electrolyte assume a $\zeta$-potential that is the same
as in~1~m\M \ KNO$_3$ ($\zeta = - 70$~mV); the calculations also
neglect the presence of dissolved CO$_2$. The later can be justified
from the bulk conductivity being dominated by the added counterions
(H$_3$0$^+$), while the former is speculative and principally
motivated by the unreasonably high $\zeta$-potentials that result when
a constant surface charge is assumed. Nevertheless, when $\phi
(\Delta_{ac} + \Delta_{nsa}) \gg 1$, which is the case for DI water
and 1~m\M \ KNO$_3$, $K^e$ is influenced by $\zeta$ in the same manner
as $M$. Therefore, while the value of $\zeta$ chosen to estimate $K^e$
influences the experimentally ascertained pore mobility $M$
[see~Eqn.(\ref{eqn:working})], it affects the theoretical prediction
of $M$ in the same manner, thereby yielding a similar best-fit value
of $\bsl$.

\subsection{Asymptotic formulas for the pore mobility} \label{sec:mobilityapprox}

In addition to the numerically exact solutions of the electrokinetic
model, let us recall an approximate analytical solution valid for
dilute composites where the $\zeta$-potential is low ($|\zeta| < \kb T
/ e$) and the double-layer thickness $\kappa^{-1}$ and Brinkman
screening length $\bsl$ are both small compared to the particle radius
$a$:
\begin{equation} \label{eqn:smoluchowski}
  U / (E \phi) = \frac{-9 \epsilon_s \epsilon_o \zeta \kappa a (\bsl /
    a)^2}{2 \eta (\kappa \bsl + 1)} \ \ (\ka \gg 1, \bsl \ll a,
  |\zeta| < \kb T / e).
\end{equation}
This formula confirms inferences drawn from numerical solutions of the
full electrokinetic model~\citep{Hill:2006b}, and, indeed,
Matos~\etal's experiments, namely that the electroosmotic flow is
inversely proportional to the particle radius. However, Matos~\etal's
experiments are characterized by $\ka < 1$ with $\bsl \sim a$. As
shown below, such conditions produce an electroosmotic flow that
scales with the reciprocal square of the particle radius at constant
$\zeta$-potential and electric field strength.

To obtain a convenient formula for small $\ka$, note that the
hydrodynamic force on an immobilized charged colloid vanishes with
respect to the electrical force when $\ka \ll 1$ and $|\zeta| < \kb T
/ e$~\citep{Hill:2006c}. Further, when $\ka \rightarrow 0$ and
$|\zeta| < \kb T / e$, the electrical force approaches the bare
Coulombe force $f^E = 4 \pi \epsilon_s \epsilon_o a \zeta
E$. Therefore, since $f^E = 4 \pi (\eta / \bsl^2) C^E
E$~\citep{Hill:2006a}, the incremental pore mobility becomes
\begin{equation} \label{eqn:huckel}
  U / (E \phi) = -3 \epsilon_s \epsilon_o \zeta (\bsl / a)^2 / \eta \
  \ (\ka \ll 1, |\zeta| < \kb T / e).
\end{equation}
It is noteworthy that the dependence of the pore mobility on particle
size is qualitatively different at high and low $\ka$. This is due to
the relationship between the surface charge density and
potential. When $\ka$ is large, the surface charge density is $\sim
\zeta \epsilon_s \epsilon_o \kappa$, so the average mobile charge
density in the hydrogel scales as $\zeta \epsilon_s \epsilon_o \kappa
\phi / a$. However, when $\ka$ is small, the surface charge density is
$\sim \zeta \epsilon_s \epsilon_o / a$, so the mobile charge density
scales as $\zeta \epsilon_s \epsilon_o \phi / a^2$. Accordingly,
balancing the electrical and Darcy drag forces gives $-E 4 \pi \zeta
\epsilon_s \epsilon_o a \phi / [(4/3) \pi a^3] = (\eta / \bsl^2) U$,
so the incremental pore mobility is $M = U / (E \phi) = -3 \epsilon_s
\epsilon_o \zeta (\bsl / a)^2 / \eta$, which agrees with
Eqn.~(\ref{eqn:huckel}).

\cite{Matos:2006} identified the electroosmotic flow in their
experiments as being inversely related to the surface area density
$\phi / a$. However, their data was insufficient to ascertain a
specific power-law dependence. While the theory clearly identifies the
mobility as having a reciprocal square dependence with particle size
at constant $\zeta$-potential when $\ka \ll 1$, and a reciprocal
dependence when $\ka \gg 1$, the theoretical interpretation of the
experiments is further complicated by the conductivity of the
composite depending on the particle size and volume fraction. In fact,
when $\ka \ll 1$, $K^e / K^\infty = 1 + O[\phi (\ka)^{-2}]$, so $U =
\phi M E$ has a weaker dependence on particle size and volume fraction
than expected if the effective conductivity were assumed
constant. More importantly, and, perhaps, surprisingly, it will be
demonstrated that ($i$) at very low bulk ionic strengths the
electroosmotic pumping velocity $U$ at constant current density ${\cal
I} = i/A$ is independent of the particle size, charge and volume
fraction; and ($ii$) the effective Darcy permeability of the polymer
skeleton evidently increases with decreasing average interparticle
surface proximity.

\begin{figure}
  \begin{center}
    \includegraphics[width=6.5cm]{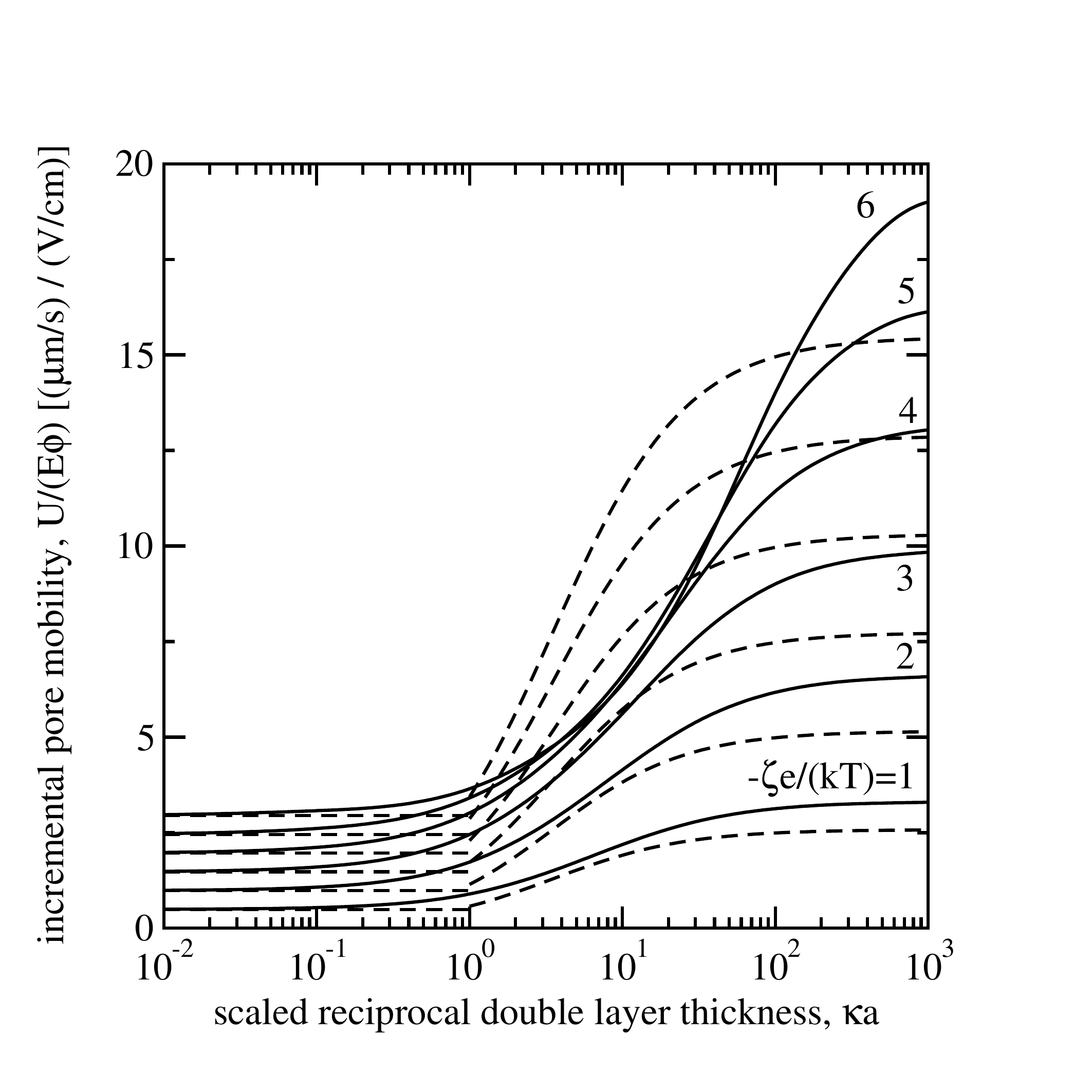}
    \includegraphics[width=6.5cm]{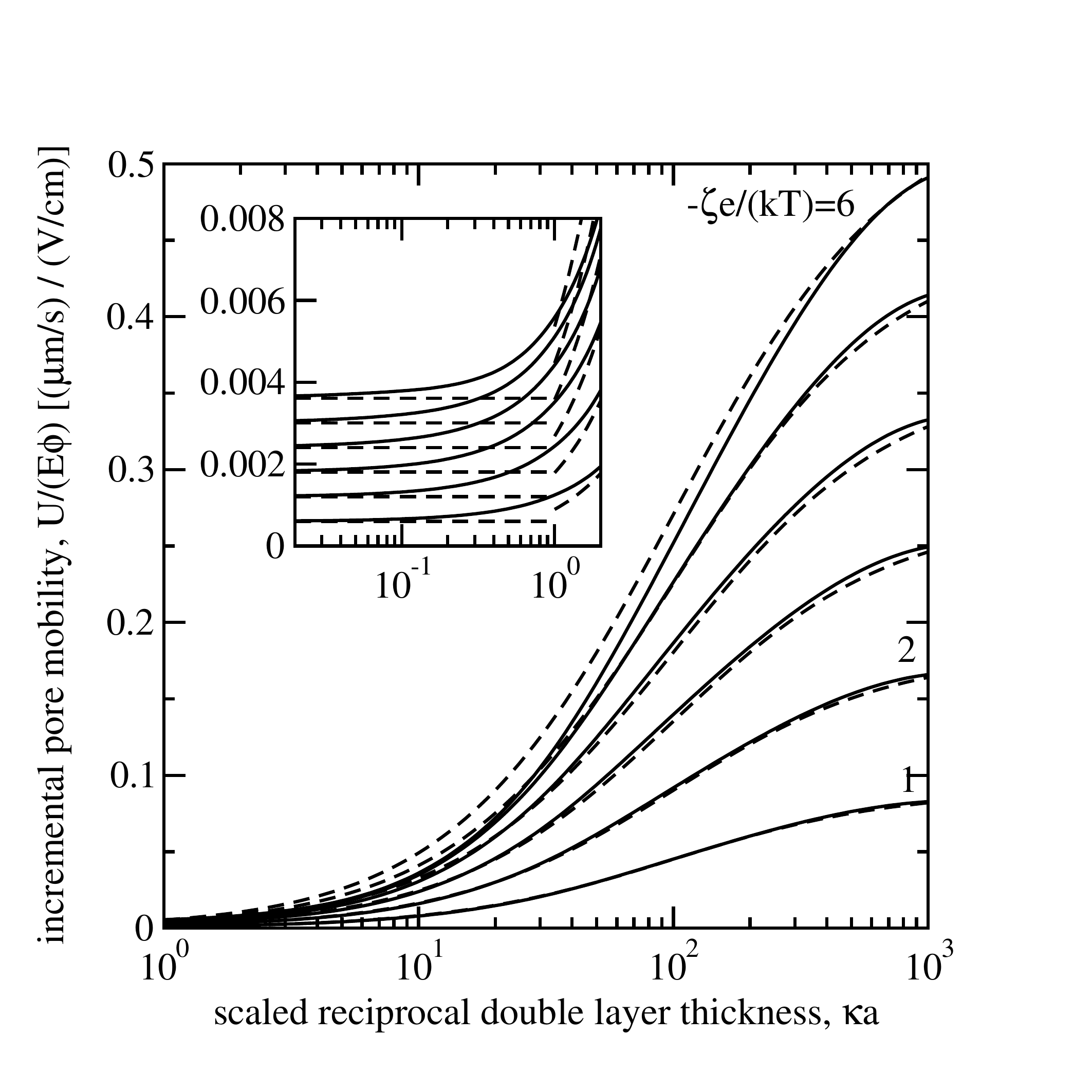}
  \end{center}
  \caption{\label{fig:mobility} The incremental pore mobility $M = U /
    (E \phi)$ as a function of the scaled reciprocal double layer
    thickness $\ka$ for various (scaled) $\zeta$-potentials $-\zeta
    e/(\kb T) = 1$, $2$, $3$, ..., $6$: aqueous KNO$_3$ at $T =
    25$\degc; $a = 3.5$ (left panel) and 100~nm (right panel); $\bsl =
    1.0$~nm. Solid lines are numerically exact solutions of the full
    electrokinetic model, and the dashed lines are approximate
    analytical solutions [Eqns.~(\ref{eqn:smoluchowski}) (large $\ka$)
    and (\ref{eqn:huckel}) (small $\ka$)].}
\end{figure}

Representative numerical calculations of the pore mobility are
compared with the asymptotic formulas in figure~\ref{fig:mobility} for
inclusions with two representative radii ($a=3.5$ and $100$~nm)
dispersed in an hydrogel with Brinkman screening length $\bsl =
1$~nm. In regions of the parameter space where the full model agrees
with Eqn.~(\ref{eqn:huckel}) (small $\ka$) and
(\ref{eqn:smoluchowski}) (large $\ka$ with $\bsl \ll a$), mobilities
for a different values of $\bsl$ can be conveniently obtained by
multiplying the ordinate by $\bsl^2$ (measured in nm$^2$).

\section{Results} \label{sec:results}

The principal methodology adopted to interpret Matos~\etal's
experiments is summarized as follows. First, the measured flux
enhancements and Eqn.~(\ref{eqn:expmobility}) are used to determine
$\Pe$. Then, from knowledge of the electrical current ($i = 3$~mA),
membrane geometry ($A = 1.64$~cm$^2$ and $L = 1.68$~mm), particle
volume fractions ($\phi = 0.02$ and 0.04) and radii ($a = 3.5$ and
$7.5$~nm), and a theoretical calculation of the effective conductivity
$K^e$, the incremental pore mobilities are
\begin{equation} \label{eqn:working}
  M = \Pe D / (\phi E L) = \Pe D K^e A / (\phi i L).
\end{equation}
With these `experimentally determined' mobilities, the electrokinetic
model provides the effective hydraulic permeabilities $\bsl^2$ of the
hydrogel in each composite. In this manner, the theoretical
interpretation, which necessitates theoretical predictions of $K^e$
and $M$, is evaluated by the consistency of the resulting Brinkman
screening lengths $\bsl$.

The electrokinetic transport calculations presented in
table~\ref{tab:conductivityapprox} were performed with unhindered ion
mobilites. However, the conductivity increment and incremental pore
mobility are practically independent of the ion mobilities, so it is
reasonable to account for hindered ion mobilities by multiplying the
unhindered bulk conductivity by a single hinderance factor. This
increases the calculated electric field strength and, therefore,
decreases the incremental pore mobility inferred from the
experimentally measured flux enhancement. In turn, smaller best-fit
values of the Brinkman screening length emerge.

The table summarizes theoretical interpretations of the five
experimental conditions for which \citeauthor{Matos:2006} reported
statistically averaged flux enhancements. Cases A--C and D--E,
respectively, identify experiments with 1~m\M \ KNO$_3$ and DI
water. Note that DI water is approximated here as an electrolyte
comprised of H$_3$O$^+$ and OH$^-$ ions, each with a concentration
$n_j^\infty = 10^{-7}$~\M, and the $\zeta$-potentials of the silica
nanoparticles are taken to be $-70$~mV in 1~m\M \ KNO$_3$ {\em and} DI
water; recall, \citeauthor{Matos:2006} reported $\zeta$-potentials
only for 1~m\M \ KNO$_3$.

The upper section of the table lists quantities that are readily
available from Matos~\etal's paper. In addition to the flux
enhancement, $J / J_0$, is a length scale $a(\phi^{-1/3} - 2)$ that
characterizes the average distance between the inclusion surfaces. The
middle section presents theoretical calculations of the polarization
and added counterion and non-specific adsorption contributions to the
conductivity increment~\citep{Saville:1983}. With the inclusion volume
fractions and bulk electrolyte conductivity $K^\infty$, these provide
the bulk conductivity of the composite $K^e$. In turn, knowledge of
the current density ${\cal I} = i / A \approx 1.83$~mA~cm$^{-2}$ (in
the composite) provides the electric field strength $E = {\cal I
}/K^e$. Note that the conductivities of the composites saturated with
DI water are comparable to those with 1~m\M \ KNO$_3$, even though the
bulk ionic strength and, hence, conductivity of the electrolyte is
four orders of magnitude lower. Clearly, particle polarization and the
added counterions play a very important role.

The lower section of the table presents calculations of the
`experimentally determined' P{\' e}clet number $\Pe = U L / D$ and
incremental pore mobility $M = U / (E \phi)$. Note that $\Pe$ may be
considered a dimensionless electroosmotic pumping velocity here, since
the membrane thickness $L$ and tracer diffusivity $D$ are same in all
cases A--E. The table presents theoretical predictions (from numerical
solutions of the full electrokinetic model) of the incremental pore
mobility for composites with a Brinkman screening length $\bsl =
1$~nm. These have been used to calculate best-fit values of $\bsl$
based on knowledge that $M$ is inversely proportional to $\bsl^2$ at
the prevailing values of $\ka < 1$ (see figure~\ref{fig:mobility}).

\begin{sidewaystable}
  \begin{center}
    \caption{\label{tab:conductivityapprox} Summary of experimental
      data~\citep[from][]{Matos:2006} and hydrogel composite
      conductivity $K^e$, mobility $M$, and permeability $\bsl^2$
      calculations with unhindered ion mobilities. Note that $\bsl_h$
      denotes Brinkman screening lengths ascertained with hindered ion
      migration (see text for details): $\zeta = -70$~mV; $i = A K^e E
      = 3$~mA; $A = 1.64$~cm$^2$; $L=1.68$~mm; $J/J_0\approx \Pe / (1
      - e^{-\Pe})$ ($\Pe > 0$).}
    
    \begin{tabular*}{\columnwidth}{@{\extracolsep{\fill}}cccccc} \hline
      case&$a$&$\phi$&$\ka$&$J/J_0$&$a(\phi^{-1/3}-2)$\\
      &(nm)& & & &(nm)\\
      \multicolumn{6}{c}{$1$~m\M \ KNO$_3$ with K$^+$ counterion}\\
      A&$3.5$&$0.02$&$0.364$&$2.36$&5.9\\
      B&$7.5$&$0.02$&$0.780$&$1.21$&12.6\\
      C&$7.5$&$0.04$&$0.780$&$2.21$&6.9\\
      \multicolumn{6}{c}{DI water (pH=7) with H$_3$O$^+$ counterion}\\
      D&$3.5$&$0.02$&$0.00364$&$1.44$&5.9\\
      E&$7.5$&$0.02$&$0.00780$&$1.19$&12.6\\
    \end{tabular*}

    \begin{tabular*}{\columnwidth}{@{\extracolsep{\fill}}ccccccc} \hline
      case&$\Delta_p$&$\Delta_{ac}$&$\Delta_{nsa}$&$K^\infty$&$K^e$&$|E|$\\
      & & & &(mS~m$^{-1}$)&(mS~m$^{-1}$)&(V~cm$^{-1}$)\\ 
      \multicolumn{7}{c}{$1$~m\M \ KNO$_3$ with K$^+$ counterion}\\
      A&$5.08$&$89.4$&$-18.4$&$14.5$&$36.6$&$5.02$\\
      B&$1.63$&$26.8$&$-8.55$&$14.5$&$20.2$&$9.06$\\
      C&$1.63$&$26.8$&$-8.55$&$14.5$&$26.0$&$7.05$\\
      \multicolumn{7}{c}{DI water (pH=7) with H$_3$O$^+$ counterion}\\
      D&$1.70 \times 10^5$&$7.90 \times 10^5$&$-1.72 \times 10^5$&$0.00549$&$86.5$&$2.12$\\
      E&$3.74 \times 10^4$&$1.73 \times 10^5$&$-3.78 \times 10^4$&$0.00549$&$18.9$&$9.70$\\
    \end{tabular*}

    \begin{tabular*}{\columnwidth}{@{\extracolsep{\fill}}ccccccc} \hline
      case&$|\Pe|$&$|M|$(exp)&$|M|(\bsl = 1.0~\mbox{nm})$~(theory)&\multicolumn{1}{c}{$\bsl = 1.0~\mbox{nm} \sqrt{\frac{|M|(\mbox{exp})}{|M|(\mbox{theory})}}$}&\multicolumn{1}{c}{$\bsl_h = \sqrt{0.50} \bsl$}&\multicolumn{1}{c}{$\bsl_h = \sqrt{0.24} \bsl$}\\
      &&$\left (\frac{\mu\mbox{m}~\mbox{s}^{-1}}{\mbox{V}~\mbox{cm}^{-1}}\right )$&$\left (\frac{\mu\mbox{m}~\mbox{s}^{-1}}{\mbox{V}~\mbox{cm}^{-1}}\right )$&\multicolumn{1}{c}{(nm)}&\multicolumn{1}{c}{(nm)}&\multicolumn{1}{c}{(nm)}\\
      \multicolumn{7}{c}{$1$~m\M \ KNO$_3$ with K$^+$ counterion}\\
      A&$2.06$&$3.91$&$1.73$&$1.50$&$1.06$&$0.74$\\
      B&$0.39$&$0.42$&$0.48$&$0.93$&$0.66$&$0.46$\\
      C&$1.87$&$1.22$&$0.48$&$1.60$&$1.13$&$0.78$\\
      \multicolumn{7}{c}{DI water (pH=7) with H$_3$O$^+$ counterion}\\
      D&$0.78$&$3.51$&$1.34$&$1.62$&$1.14$&$0.79$\\
      E&$0.36$&$0.35$&$0.29$&$1.09$&$0.77$&$0.53$\\
\hline
    \end{tabular*}
  \end{center}
\end{sidewaystable}

The scenario that emerges from these calculations is, perhaps, more
complex than suggested by Matos~\etal's qualitative interpretation of
the data. For example, comparing the P{\' e}clet numbers for cases A
and B, where the particle size is doubled at a fixed inclusion volume
fraction, suggests that $U$ scales as $a^{-2}$. While this is
consistent with theoretical expectations if $\bsl$ is a constant,
doubling the particle size decreases the mobility by almost an order
of magnitude. The anomaly can be reconciled by allowing $\bsl$ to
vary, but it is not clear that the resulting best-fit values of $\bsl$
are acceptable. A similar discussion applies to the comparison of
cases D and E (with DI water). Next, comparing cases B and C, where
the inclusion volume fraction is doubled with a fixed particle size,
the P{\' e}clet numbers indicate that $U$ more than quadruples,
whereas the theory requires $U$ to double. Furthermore, the mobility
triples, while the theory demands a constant mobility if $\bsl$ is
constant.

\begin{figure}
  \begin{center}
    \includegraphics[width=8cm]{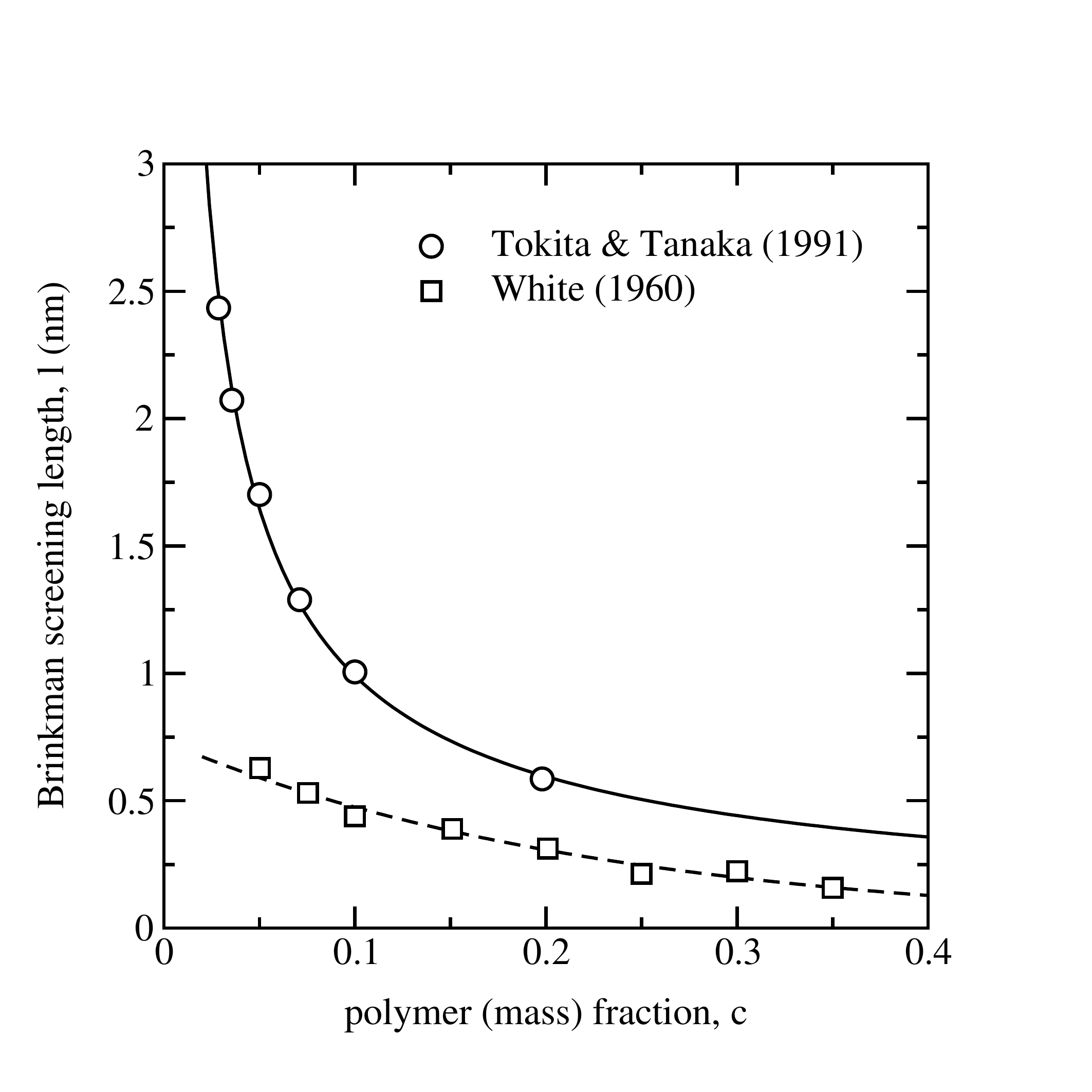}
  \end{center}
  \caption{\label{fig:permeability} The Brinkman screening length
    $\bsl$ (nm) of polyacrylamide hydrogels, ascertained from
    pressure-driven flow experiments in thin membranes, as a function
    of the polymer (mass) fraction $c$: \cite{White:1960} (squares)
    and \cite{Tokita:1991} (circles). Curves are empirical fits to the
    data: $\bsl = 0.182 c^{-0.735}$ (nm) (solid); $\bsl = 0.735
    \exp{(-4.36 c)}$ (nm) (dashed).}
\end{figure}

Figure~\ref{fig:permeability} compares the Brinkman screening lengths
ascertained from independent pressure-driven flows in thin
polyacrylamide gel membranes without nano-inclusions. The two sets of
data reported by \cite{White:1960} and \cite{Tokita:1991} are clearly
very different, and it is unfortunate, perhaps, that the later authors
did not reconcile these discrepancies. Nevertheless, in favor of their
results is the consistency of their data with scaling
theory. Accordingly, a semi-empirical fit to their data,
\begin{equation} \label{eqn:tokita}
  \bsl \approx 0.182 c^{-0.735} \sim c^{-3/4} \ \ (\mbox{nm}),
\end{equation}
gives $\bsl \approx 0.8$~nm at the polymer concentration $c = 0.13$
reported by \cite{Matos:2006}. This is clearly smaller than all the
values of $\bsl$ (spanning the range $0.93$--$1.62$~nm) ascertained
from the theoretical interpretation of Matos~\etal's data presented in
table~\ref{tab:conductivityapprox}.

Note, with a fixed value of $\bsl = 0.8$~nm, table~\ref{tab:nktheory}
presents theoretical calculations with parameters corresponding to
cases A--E in table~\ref{tab:conductivityapprox}. Again, all the
calculations (leading to the incremental pore mobility and electrical
condictivity) have been performed with unhindered electrolyte ion
mobilities. The results in the upper and lower sections of the table
are, respectively, with a 1~m\M \ KNO$_3$ electrolyte (with K$^+$
counterions) and DI water (with H$_3$O$^+$ counterions).


\begin{sidewaystable}
  \begin{center}
    \caption{\label{tab:nktheory} Theoretical expectations for
      Matos~\etal's experiments. Calculations with 1~m\M \ KNO$_3$
      supporting electrolyte (upper section) and DI water (lower
      section), both with unhindered ion mobilities: $\zeta = -70$~mV;
      $\bsl = 0.8$~nm; $i = A K^e E = 3$~mA; $A = 1.64$~cm$^2$;
      $L=1.68$~mm; $J/J_0\approx \Pe / (1-e^{-\Pe})$ ($\Pe > 0$).}
    \begin{tabular*}{\columnwidth}{@{\extracolsep{\fill}}cccccccccccc} \hline
      $a$&$\phi$&$\ka$&$\Delta_{p}$&$\Delta_{ac}$&$\Delta_{nsa}$&$K^\infty$&$K^e$&$M$&$E$&$\Pe$&$J/J_0$\\
      (nm)&&&&&&(mS~m$^{-1}$)&(mS~m$^{-1}$)&$\left (\frac{\mu\mbox{m}~\mbox{s}^{-1}}{\mbox{V}~\mbox{cm}^{-1}}\right )$&(V~cm$^{-1}$)\\
      \multicolumn{12}{c}{$1$~m\M \ KNO$_3$ with K$^+$ counterion}\\
      $3.5$&$0.02$&$0.364$&$5.08$&$89.4$&$-18.4$&$14.5$&$36.6$&$1.11$&$5.02$&$0.57$&$1.31$\\
      $7.5$&$0.02$&$0.780$&$1.63$&$26.8$&$-8.55$&$14.5$&$20.2$&$0.307$&$9.06$&$0.28$&$1.15$\\
      $7.5$&$0.04$&$0.780$&$1.63$&$26.8$&$-8.55$&$14.5$&$26.0$&$0.307$&$7.05$&$0.44$&$1.24$\\
      \multicolumn{12}{c}{DI water (pH=7) with H$_3$O$^+$ counterion}\\
      $3.5$&$0.02$&$0.00364$&$1.70\times10^5$&$7.90\times10^5$&$-1.72\times10^5$&$0.00549$&$86.5$&$0.860$&$2.12$&$0.19$&$1.10$\\
      $7.5$&$0.02$&$0.00780$&$4.29\times10^4$&$1.98\times10^5$&$-4.34\times10^4$&$0.00549$&$18.9$&$0.188$&$9.70$&$0.19$&$1.10$\\
      $7.5$&$0.04$&$0.00780$&$4.29\times10^4$&$1.98\times10^5$&$-4.34\times10^4$&$0.00549$&$37.8$&$0.189$&$4.85$&$0.19$&$1.10$\\
      \hline
    \end{tabular*}
  \end{center}
\end{sidewaystable}

The calculations with a KNO$_3$ electrolyte indicate that the
mobilities, P{\' e}clet numbers, and flux enhancements exhibit similar
qualitative trends as the experiments. Howevever, the quantitative
trends are poor. Note that the conductivity increments are practically
the same as those presented in table~\ref{tab:conductivityapprox} with
$\bsl = 1$~nm. This reflects the extremely weak role of microscale
convection in polarizing the inclusions, and, of course, that added
counterions and non-specific adsorption are an $O(\phi)$ contribution
to equilibrium.

The calculations in the lower section of the table, with DI water as
the electrolyte, demonstrate that the electroosmotic flow velocity, as
indicated by $\Pe$ or $J/J_0$, is practically independent of the
particle size and volume fraction at sufficiently low bulk ionic
strengths. If the discrepancy between the electroosmotic flow velocity
$U = \Pe D / L \approx 37$~nm~s$^{-1}$ and the experimentaly
determined value $U \approx 74$~nm~s$^{-1}$ for DI water (when $a
(\phi^{-1/3}-2) \approx 11.8$) can be attributed to the neglect of
hindered ion migration, which leads to an incorrect estimate of the
bulk conductivity $K^\infty$, then the effective electrolyte ion
hinderance factor is suggested to be about $37 / 74 \approx
0.50$. This is in reasonable agreement with independent measurements
of the hindered diffusion coefficients of KCl and D$_2$O (and two
larger molecules) in polyacrylamide hydrogels. In particular, at the
bulk polymer concentration used in Matos~\etal's experiments ($c
\approx 0.13$), \cite{White:1961} measured diffusion coefficients of
KCl and D$_2$O that are factors of $1.2/1.8 \approx 0.67$ and $1.2 /
2.2 \approx 0.55$ smaller than their respective unhindered values.

Accordingly, if it is assumed that the electrical conductivity of the
membranes is indeed reduced by a factor of 0.50, and, furthermore, the
mobilities of all electrolyte ions are reduced by the same factor,
then the incremental pore mobilities $M$ inferred from Matos~\etal's
experiments should be a factor 0.50 smaller than the values listed in
table~\ref{tab:conductivityapprox}. Moreover, the best-fit Brinkman
screening lengths $\bsl$ in the table should be reduced by a factor
$\sqrt{0.50} \approx 0.71$. These values are denoted $\bsl_h =
\sqrt{0.50} \bsl$ in the lower section of
table~\ref{tab:conductivityapprox}. It is noteworthy that their range
$0.66$--$1.14$~nm is much more representative of the values expected
from the pressure-driven flows of \cite{White:1960} and
\cite{Tokita:1991}.

Knowledge of the electrical conductivity of the electrolyte-saturated
hydrogel would provide an independent quantitative measure of the
influence that the polymer has on electrolyte ion mobilities while
under the influence of an electric field.  Unfortunately, the
electrical conductiviy of the hydrogels in Matos~\etal's experiments
is unknown. However, data in Matos's
thesis~\citep[fig.~4.29][]{Matosthesis} indicates that the voltage
drop across the entire cell (\ie, between the electrodes) when $i =
3$~mA is about $12$~V when gels without inclusions are saturated with
1~m\M KNO$_3$. If this voltage is assumed to reflect only the combined
electrical resistances of the membrane and electrolyte sandwiched
between the electrodes, then the conductivity of the membrane is
\begin{equation}
  K^e = K^\infty / [K^\infty A \Delta / (i L)-(L'/L)(A / A')].
\end{equation}
Here, $\Delta \approx 12$~V is the total potential drop, $K^\infty
\approx 14.5$~mS~m$^{-1}$ is the conductivity of 1~m\M KNO$_3$
electrolyte (from knowledge of the limiting conductances of $K^+$ and
NO$_3^-$), $L' = 6-1.68 = 4.32$~mm is twice the distance between the
membrane and one electrode, and $A' \approx 2.84$~cm$^2$ is the clear
cross-sectional area between the membrane and electrodes. It follows
that $K^e \approx K^\infty / (5.66 - 1.56) \approx 0.24
K^\infty$. While this indicates that the polymer significantly hinders
ion migration, the calculation is likely to over-estimate the
hindrance, because it neglects the (unknown) potential drop due to
electrochemical reactions.

Nevertheless, with a hinderance factor of $0.24$ adoped to provide a
lower-bound on the membrane electrical conductivity, the corresponding
Brinkman screening lengths, denoted $\bsl_h = \sqrt{0.24} \bsl$ in the
lower section of table~\ref{tab:conductivityapprox}, are a factor
$\sqrt{0.24}\approx 0.49$ smaller than the best-fit values ascertained
with unhindered ion mobilities. The range $\bsl_h = 0.53$--$0.79$~nm
is representative of the values expected from the pressure-driven
flows of \cite{White:1960} and \cite{Tokita:1991}, but, as highlighted
below, such a favorable comparision may be misleading.

The analysis summarized in table~\ref{tab:conductivityapprox} has led
to an upper range $\bsl \approx 0.9$--1.60~nm ascertained with
unhindered ion mobilities, and a lower range $\bsl_h \approx
0.53$--$0.79$~nm ascertained using a lower-bound estimate of hindered
ion mobilities. Note that fluctuations within each range correspond to
unreasonably large fluctuations in the bulk polymer density, and such
fluctuations cannot be reconciled by osmotic swelling. It therefore
remains to elucidate these significant variations in gel permeability.

\section{Discussion} \label{sec:discussion}

\begin{figure}
  \begin{center}
    \includegraphics[width=8cm]{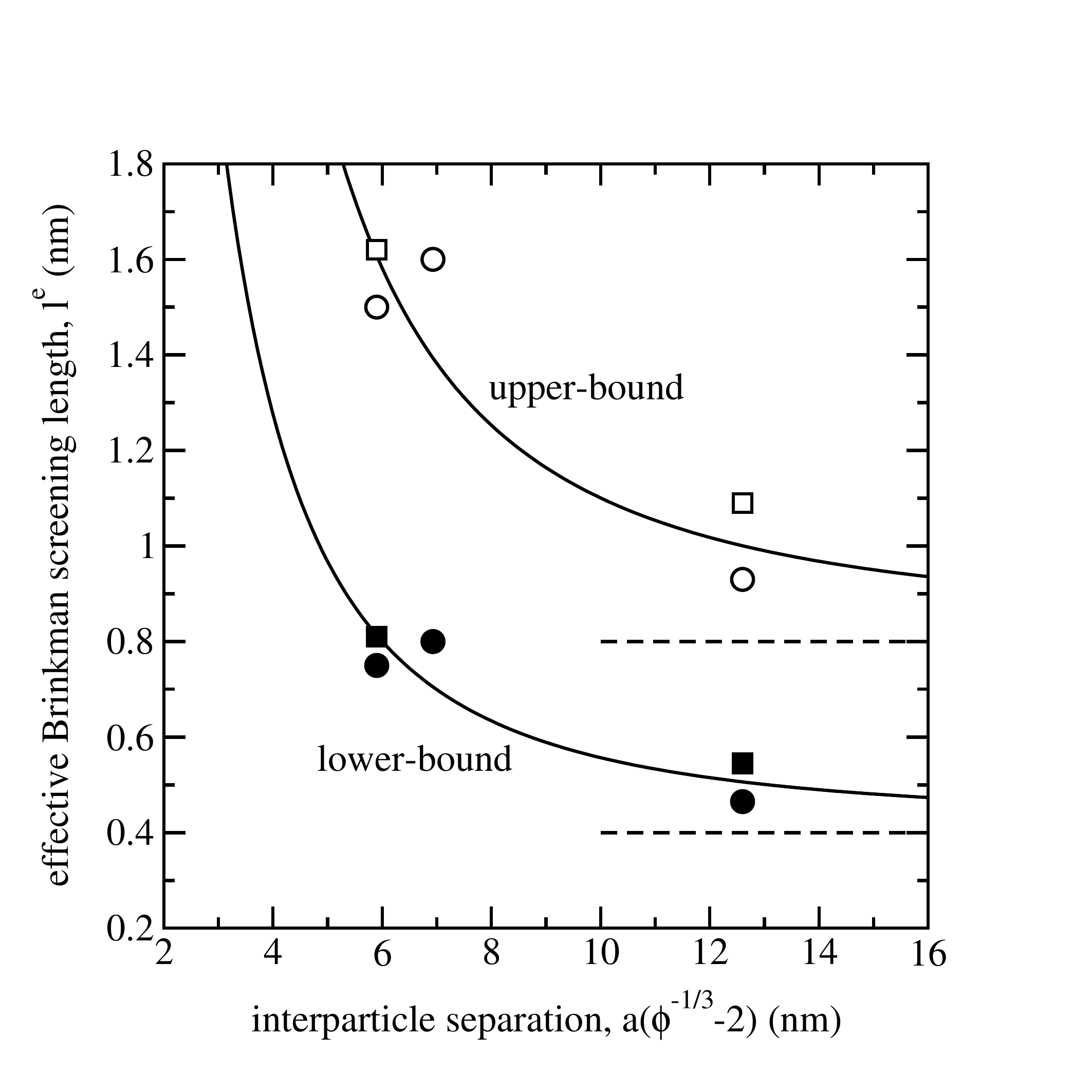}
  \end{center}
  \caption{\label{fig:convergence} The effective Brinkman screening
    length $\bsl^e$ as a function of the average distance between the
    surfaces of nearest neighbor inclusions in random dispersions $a
    (\phi^{-1/3} -2)$: upper-bound estimates based on theoretical
    interpretation of experiments with unhindered ion mobilities (open
    symbols); lower-bound estimates achieved with a lower-bound
    estimate of hydrogel electrical conductivity (filled symbols). See
    text for details. Circles and squares denote, respectively, cases
    A--C with 1~m\M KNO$_3$ electrolyte, and cases D and E with DI
    water (see table~\ref{tab:conductivityapprox}). The solid lines
    are empirical best-fit (quadratic) decays: $\bsl^e = \bsl [1 +
    47.3 / \{(\phi^{-1/3} -2) a / \bsl \}^2]$ with $\bsl = 0.83$~nm
    (upper curve); $\bsl^e = \bsl [1 + 184 / \{(\phi^{-1/3} -2)a /
    \bsl \}^2]$ with $\bsl = 0.42$~nm (lower curve). Dashed lines
    indicate the values of $\bsl$ expected from pressure-driven flows
    in polyacrylamide gels with a polymer concentration $c \approx
    0.13$: $\bsl \approx 0.8$~nm \citep{Tokita:1991} (upper line);
    $\bsl \approx 0.4$~nm \citep{White:1960} (lower line).}
\end{figure}

One clue to resolving the apparent shortcomings is to recognize that
two distinctly different values of the best-fit Brinkman screening
lengths emerge from the experiments. Considering, for example, the
values ascertained with unhindered ion mobilities, cases B and E yield
values of $\bsl$ close to 1~nm, whereas cases A, C and D yield values
close to 1.5~nm. These differences do not correlate with the
electrolyte, particle size or volume fraction alone. However, there is
strong correlation with a length scale that characterizes the average
distance between closest neighbors in a uniform random
dispersion. Letting the characteristic center-to-center distance be $a
\phi^{-1/3}$, the distance between the surfaces is $a (\phi^{-1/3} -
2)$. This measure is listed in the last column of the upper section of
table~\ref{tab:conductivityapprox}, and its relationship to the
effective permeability is plotted in figure~\ref{fig:convergence}.

Recall, the best-fit values of $\bsl$, ascertained without hindered
ion mobilities influencing the membrane electrical conductivity,
provide an upper-bound on the effective gel permeability (open
symbols). Furthermore, the lower-bound on the electrical conductivity,
established above from the voltage-current relationship for the cell,
provides a lower-bound-estimate of the gel permeability (filled
symbols). As demonstrated in the figure~\ref{fig:convergence}, these
upper- and lower-bounds approach values expected from independent
pressure-driven flow experiments in homogeneous (\ie, without
inclusions) polyacrylamide
gels~\citep{Tokita:1991,White:1960}. Moreover, the approach occurs as
the separation $a (\phi^{-1/3}-2)$ becomes large compared to $\bsl$.

This quantitative analysis supports Matos~\etal's suggestion that
particle interactions play a necessary role in endowing the composites
with a significant electroosmotic pumping capacity.  Nevertheless, the
physical origin of such interactions and, indeed, their quantitative
influence on the electroosmotic pumping capacity {\em and} bulk
conductivity remain unknown.

\citeauthor{Matos:2006} suggested that electroosmotic flow creates
pathways that somehow link particles in close proximity, as expected
in a percolating network. However, their neutron scattering data did
not reflect fractal-like correlations. Another possibility is that
polymer structure (\eg, partial chain orientation and/or cross-linking
density) is altered near the particle surfaces, producing shells of
hydrogel that are much more electrically conductive (\ie, with higher
ion permeabilities) and hydrodynamically permeable than in the
bulk. Swelling after the synthesis could also generate shells of void
space.

It is tempting to draw upon Matos~\etal's data with $a=3.5$~nm and
$\phi = 0.005$, 0.01 and 0.2 \citep[as seen in fig.~8
of][]{Matos:2006} to help elucidate the role of particle volume
fraction and size.  However, it is evident from these single
experiments, and the averaged data (used exclusively above), that
there are significant statistical fluctuations from one experiment to
another.  For example, the flux enhancements from the experiments
presented in figures~1 and 8 of Matos~\etal's paper with DI water
(with $a=3.5$~nm and $\phi = 0.02$) are, respectively, $\approx 1.58$
and 1.63, whereas the flux enhancement from repeated measurements
under the same conditions was reported as $1.44 \pm 0.01$ (as adopted
in table~\ref{tab:conductivityapprox} above). It is also difficult to
draw firm conclusions from the experiments with DI water, since the
$\zeta$-potentials are unknown, further hampering theoretical
estimates of the bulk conductivity and incremental pore mobility.

Finally, note that silica nanoinclusions have been demonstrated to
enhance the permeability of glassy polymeric matrices to
gases~\citep{Merkel:2002}. This unexpected behavior is attributed to
an overall increase in the free volume of the polymer. Moreover,
recent theoretical studies show that the increase in free volume is
localized to the particle-polymer interface~\citep{Hill:2006e}. The
interpretation of Matos~\etal's experiments summarized in
figure~\ref{fig:convergence} suggests that a similar influence
prevails when silica nanoparticles are immobilized in swollen
water-saturated polymer networks.

\section{Summary} \label{sec:summary}

The full electrokinetic transport model proposed by \cite{Hill:2006b}
was used to interpret recent experiments where \cite{Matos:2006}
immobilized silica nanospheres in an uncharged polayacrylamide
hydrogel matrices and applied an electric field to electroosmotically
enhance the otherwise a purely diffusive flux of an uncharged tracer
across the composite membrane.

Convenient simplifications of the full model, leading to simple
closed-form formulas, were derived for the limit where the bulk ionic
strength is low and the inclusion radius is small; these complement
earlier analytical solutions for situations where the Debye and
Brinkman screening lengths are small compared to the particle
radius. The analytical results were demonstrated to yield accurate
approximations of the incremental pore mobility from numerically exact
solutions of the full electrokinetic model. Figure~\ref{fig:mobility}
identifies practical bounds on the parameter space where these
convenient approximations are accurate.

A simple (quasi-steady) convective-diffusion model for the tracer flux
was derived to quantitatively interpret the flux-enhancement
diagnostic reported by \citeauthor{Matos:2006} The general solution
depends on the (changing) concentration of tracer in the source and
sink reservoirs, but it was assumed here that the concentration in the
sink was small enough to permit an approximation that relates the
scaled electroosmotic flow velocity ($\Pe = U L / D > 0$) directly to
the flux enhancement $J / J_0 > 1$.

Several approaches were taken to compare the theory and
experiments. One involved selecting the Brinkman screening length of
the polymer skeleton as a fitting parameter and using the experiments
and theory to establish values that yield `correct' flux
enhancements. With unhindered electrolyte migration, the effective
Brinkman screening length of the hydrogels was found to span the range
$0.93$--$1.63$~nm, depending, foremost, on the average particle
separation, as measured by $a (\phi^{-1/3} - 2)$. Evidently, as the
particle separation diverges, the effective Brinkman screening length
approaches a value ($\approx 0.8$~nm) that is expected from
independent pressure-driven flow experiments by~\cite{Tokita:1991}. On
the other hand, allowing for hindered electrolyte ion migration, a
lower-bound-estimate of the efffective Brinkman screening length gives
a range $0.46$--$0.79$~nm. Similarly, as the particle separation
diverges, the effective Brinkman screening length approaches a value
($\approx 0.4$~nm) that is expected from independent pressure-driven
flow experiments by~\cite{White:1960}.

Several means of reconciling the apparent differences between theory
and experiment were proposed, all of which require more extensive and
complete experimental investigation to draw firm conclusions. Firstly,
this analysis supports Matos~\etal's suggestion that particle
interactions, which are neglected entirely in the present theory (due
to the small inclusion volume fraction), significantly increase the
electroosmotic pumping capacity of the composites. A model to quantify
the influence of such interactions has not been pursued, and it is not
clear that the enhanced permeability arises from particle-particle or
particle-polymer interactions, or, perhaps, a combination of both.

While the electroosmotic pumping capacity of a composite is
independent of the bulk electrical conductivity, when the electric
field driving the flow is known, in practice it is the electrical
current that is known accurately. This necessitates an accurate
experimental determination of the bulk conductivity to complement a
satisfactory theoretical interpretation of experiments, and thus
permit an accurate prediction of the pumping capacity as a function of
current density.

Clearly, it would be advantageous if future experimental efforts
reported the conductivity of hydrogels with and without inclusions,
preferentially when uniformly saturated with well-characterized
electrolytes, including DI water. Such diagnostics should be performed
using a high-frequency electric field while the membrane is in its
steady or quasi-steady state, \ie, when subjected to a steady electric
field. Note, however, that care must be taken to separate the
influences of the bulk and near-electrode
regions~\citep{Hollingsworth:2003}.

  Supported by the Natural Sciences and Engineering Research Council
  of Canada (NSERC), through grant number 204542, and the Canada
  Research Chairs program (Tier II). 

  
\bibliography{/home/rhill/latex/bibliographies/global}
\bibliographystyle{elsart-harv.bst}


\end{document}